\documentclass[amssymb,amsmath,prd,twocolumn,superscriptaddress,nofootinbib]{revtex4-1}

\usepackage{graphicx}
\usepackage{bm}
\usepackage{amssymb,amsmath}
\usepackage{mathrsfs}
\usepackage{latexsym}
\usepackage{color}
\usepackage[normalem]{ulem}
\usepackage{dcolumn}
\usepackage[colorlinks=true,citecolor=blue,urlcolor=blue]{hyperref}
\usepackage[usenames,dvipsnames]{xcolor}
\usepackage[caption=false]{subfig}

\newcommand{\beq}{\begin{equation}}
\newcommand{\eeq}{\end{equation}}
\newcommand{\ba}{\begin{array}}
\newcommand{\ea}{\end{array}}
\newcommand{\bea}{\begin{eqnarray}}
\newcommand{\eea}{\end{eqnarray}}
\newcommand{\bc}{\begin{center}}
\newcommand{\ec}{\end{center}}

\newcommand{\Msolar}{{\rm M}_{\odot}}
\newcommand{\Mmax}{M_{\rm max}}
\newcommand{\Mmin}{M_{\rm min}}
\newcommand{\Mchirp}{{\mathcal M}}
\newcommand{\chieff}{\chi_{\mathrm{eff}}}

\allowdisplaybreaks

\newcommand{\CCA}{\affiliation{Center for Computational Astrophysics, Flatiron Institute, 162 5th Ave, New York, NY 10010, United States}}
\newcommand{\StonyBrook}{\affiliation{Department of Physics and Astronomy, Stony Brook University, Stony Brook NY 11794, United States}}

\definecolor{azgreen}{rgb}{0.03,0.47,0.19}
\definecolor{kcmagenta}{rgb}{0.54, 0.17, 0.88}
\definecolor{chorange}{rgb}{0.851, 0.372, 0.007}

\newcommand{\nn}{\nonumber}

\begin{document}

\title{Inferring the maximum and minimum mass of merging neutron stars with gravitational waves}

\author{Katerina Chatziioannou}\CCA
\author{Will M. Farr}\CCA \StonyBrook

\date{\today}

\begin{abstract}
We show that the maximum and the minimum mass of merging neutron stars can be estimated with upcoming gravitational wave observations.
We simulate populations of binary neutron star signals and model their mass distribution including upper and lower cutoffs.
The lower(upper)  limit can be measured to $\sim 0.2(0.1)\Msolar$ with $50$ detections if the mass distribution supports neutron stars
with masses close to the cutoffs.
The upper mass limit informs about the high-density properties of the neutron star equation of state, while the lower limit signals the divide between
neutron stars and white dwarfs.
\end{abstract}

\maketitle

\section{Introduction}
\label{sec:intro}

Despite being first detected more than $50$ years ago, the properties of neutron stars (NSs) such as their possible masses and sizes are
still uncertain~\cite{Ozel:2016oaf,Lattimer:2015nhk}. Coalescences of NSs, now observable with gravitational waves
(GWs)~\cite{TheLIGOScientific:2017qsa,Abbott:2020uma} by LIGO~\cite{TheLIGOScientific:2014jea} and Virgo~\cite{TheVirgo:2014hva},
can offer information about both, through measurement of the binary component masses
and tidal interactions between the two stars as they are about to merge~\cite{Flanagan:2007ix}. The masses of NSs offer information both about the astrophysics
of compact objects, and about the dense matter NSs are made of.

Stable nonrotating NSs have a maximum possible gravitational mass $\Mmax$ beyond which internal pressure cannot support them against gravitational collapse toward
black holes (BHs). The maximum mass is a function of the unknown equation of state (EoS) of NSs that governs the properties and composition of
their interiors though rotation can offer additional support, increasing $\Mmax$ by about 20\%~\cite{Lasota:1995eu}.
The detection of heavy pulsars
through radio surveys has placed a robust lower limit of
$\Mmax\gtrsim2\Msolar$~\cite{Antoniadis:2013pzd,Cromartie:2019kug}, suggesting that the high-density EoS is stiff enough to
support them against collapse. This poses a challenge in particular for models predicting phase transitions inside NSs that result in a softening
of the EoS and lower the maximum mass possible~\cite{Han:2018mtj}.
The mass distribution of galactic NSs offers tentative evidence for an upper cutoff at $2.0-2.6\Msolar$~\cite{Antoniadis:2016hxz,Alsing:2017bbc}, while assuming
that merging NSs follow the galactic double NS distribution and produce the observed gamma ray bursts
led to  $\Mmax \lesssim 2.0-2.2\Msolar$ before the detection of GWs~\cite{Lawrence:2015oka}.

Current GW observations are consistent with NSs with masses below $2\Msolar$, but they have been used to study the maximum NS mass by
considering the merger outcome~\cite{Bauswein2013,Margalit:2019dpi,Bauswein:2020aag} or EoS modeling.
Interpreting the electromagnetic counterpart to GW170817 as supporting the formation of a hypermassive
NS remnant that eventually collapsed to a BH and assumptions about the post-merger evolution of the system suggest $\Mmax\lesssim 2.3\Msolar$~\cite{Margalit:2017dij,Ruiz:2017due,Shibata:2017xdx,2018ApJ...852L..25R,Shibata:2019ctb,LIGOScientific:2019eut}.
In parallel, tidal interactions in GW170817 offer constraints on the low-density EoS. Extrapolating to high densities using a model for the EoS based on a
gaussian process conditioned on existing nuclear models yields $\Mmax\lesssim 2.4\Msolar$~\cite{Landry:2018prl,Essick:2019ldf}.

It is unknown whether stellar evolution can produce NSs up to the maximum mass
allowed by nuclear physics and BHs down to the most massive NSs.
X-ray observations provide tentative evidence for a mass gap between the heaviest NS and the lightest BH,
though its existence is under debate~\cite{2011ApJ...741..103F,Kreidberg_2012}. Recent observations
suggest the existence of a $2.6-6.2M_{\odot}$ compact object~\cite{Thompson:2018ycv}, though this conclusion 
is under debate~\cite{vandenHeuvel:2020chh,Thompson:2020nbd}. The secondary component of 
GW190814 has a mass of $2.5-2.7$, but it remains unclear if this is a NS or a BH~\cite{Abbott:2020khf}. 
The minimum mass of astrophysical NSs $\Mmin$ is expected to be entirely driven by their formation
mechanism and might inform the divide between NSs and the next most-compact object, white dwarfs (WDs).

Observational campaigns and improved detector sensitivity are expected to yield dozens of binary NS (BNS) detections through GWs in the coming years~\cite{Aasi:2013wya}.
We examine whether these observations can be used to extract the mass distribution of coalescing NSs and in particular the maximum and minimum mass.
We find that $\Mmax$ can me measured to within $\sim 0.2M_{\odot}$ and $\Mmin$ to within $\sim 0.1M_{\odot}$ at the 90\% level with $50$ observations if the
mass distribution has support for heavy and light NSs.
The $\Mmax$ constraint can reduce the uncertainty about the pressure at 4.5 times saturation density by $\sim20$\%.
However, if binary formation mechanisms lead to a mass distribution that smoothly tails off on the high or low end,
the measurement uncertainties for $\Mmax$ and $\Mmin$ correspondingly increase.
Our estimates are conservative as we impose no restrictions on the potential NS spins, which leads to larger mass uncertainties
compared to assuming that merging NSs are slowly spinning, per galactic observations~\cite{Tauris:2017omb}.
Our mass estimates are solely based on the inferred masses and are not subject to systematics related to tidal inference, EoS modeling, or the
interpretation of a possible electromagnetic counterpart.

\section{A population of BNS signals}
\label{sec:popSimandModel}

Similar to BH binaries, the mass distribution of NSs in binaries depends on the formation mechanism. The minimum possible NS mass
 is related to the transition between WDs and NSs, while the absolute maximum is driven by the unknown EoS. However, it is unclear if binary evolution can result
 in systems with components close to the extremes, either for BHs or NSs.
 Given these uncertainties and
 evidence suggesting that merging NSs have a different mass distribution than the observed
galactic double NSs~\cite{Abbott:2020uma}, we consider three mass distributions and simulate populations of potentially observable BNSs:
(i) the masses $m_1,m_2$ are uniformly distributed in $[\Mmin=1,\Mmax=\{2.0,2.2\}]\Msolar$ with $m_1>m_2$ (``Uniform"),
(ii) the primary mass $m_1$ is uniformly distributed in $[\Mmin=1,\Mmax=\{2.0,2.2\}]\Msolar$ while the mass ratio $q\equiv m_2/m_1$ favors
equal masses as suggested by~\cite{Dominik_2012}; we use a $q^{3}$ distribution (``UniformQ"), and
(iii) the primary mass $m_1$ is distributed according to a bimodal distribution as suggested in~\cite{Antoniadis:2016hxz,Alsing:2017bbc} based on galactic NSs
with $\Mmin=1\Msolar,\Mmax=\{2.0,2.2\}\Msolar$ while the mass ratio goes as $q^{3}$.
We repeat the analysis of~\cite{Alsing:2017bbc} including the recent observation
from~\cite{Cromartie:2019kug} and select a fair draw from the mass distribution posterior (``Bimodal")\footnote{Model and samples are available in
\href{https://github.com/farr/AlsingNSMassReplication}{https://github.com/farr/AlsingNSMassReplication}.}.
We do not consider a single gaussian distribution~\cite{Farrow:2019xnc}, as GW190425 might suggest that merging BNSs
do not follow it~\cite{Abbott:2020uma}.
In all cases, the sharp upper cutoff in the mass distribution is related to the NS EoS, however, the ``Bimodal" distribution exemplifies a situation where  
the binary formation mechanism reduces the rate of heavy NSs in binaries independently of the EoS.

Given the parameters of a simulated BNS, we approximate
 measurement uncertainty with the methods described in Appendix~\ref{sim}. Rather than assuming
a gaussian likelihood in the parameters of interest as is common, we utilize the fact that information about the binary parameters comes from modeling the
phase evolution of the GW signal~\cite{Ng:2018neg}.
We instead assume gaussian likelihoods in the coefficients of the Taylor expansion of the GW phase around small velocities. This
method is able to capture the effect of the mass-spin degeneracy~\cite{PhysRevD.49.2658}, resulting in asymmetric likelihoods for the mass ratio and the effective
spin of the binary. In order to be conservative, we do not impose that the spin of the NSs is small, which results in a larger uncertainty on the binary mass ratio, see
for example the high-spin and low-spin inference for GW170817 in~\cite{Abbott:2018wiz}.

We model the simulated BNS populations with the hierarchical formalism of~\cite{Mandel:2009pc} while simultaneously fitting for the true masses of each observed
event~\cite{Hogg_2010}. We consider two mass models for the primary mass and the binary mass ratio: (i) a power law for both -with which we fit the
``Uniform" and ``UniformQ" populations-
\begin{equation}
P(m_1,q|\alpha,\beta)\sim m_1^{-\alpha}q^{\beta},
\end{equation}
and (ii) a two-gaussian distribution for the primary mass with a power law for the mass ratio
-with which we fit the ``Bimodal" population-
\begin{align}
&P(m_1,q|A,\mu_1,\sigma_1,\mu_2,\sigma_2)\sim\nn \\
& \left[A \kappa_1 {\cal{N}}(m_1;\mu_1,\sigma_1)+(1-A) \kappa_2 {\cal{N}}(m_1;\mu_2,\sigma_2)\right]q^{\beta}.
\end{align}
where the $\kappa_{1,2}$ are chosen so each Gaussian integrates to 1 over $\Mmin
< m_1 < \Mmax$, which means $A$ is the fraction of NSs associated to the first
Gaussian.

GW observations are subject to a strong selection bias toward more massive events that emit stronger signals. For low-mass binaries the selection effect can be
analytically approximated as the probability that an event is observed is proportional to $\Mchirp^{5/2}$. We take this selection effect into account both in our simulated
population (where the observed population contains more heavy systems than the intrinsic population) and in the hierarchical inference in order to avoid
 biases~\cite{2004AIPC..735..195L,Mandel:2018mve}.

\begin{figure}[h]
\includegraphics[width=\columnwidth,clip=true]{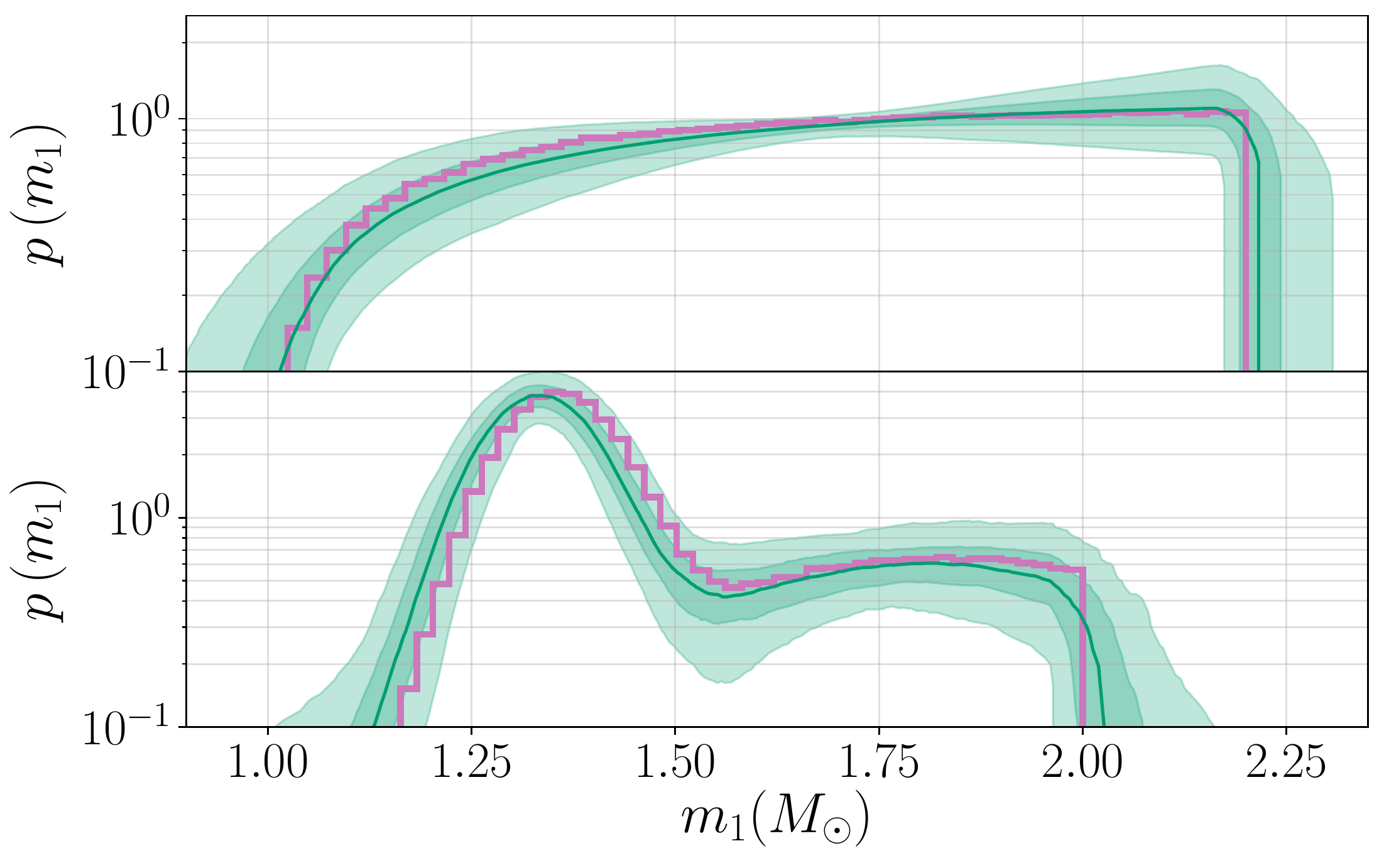}
\caption{Primary mass distribution inferred from $50$ simulated BNS detections. The pink histogram is the true distribution.
The green line and shaded regions are the median, 50\%, and 90\% credible intervals of the inferred distribution respectively. Top: uniform distribution.
Bottom: bimodal distribution.
}
\label{fig:Mpop}
\end{figure}

Figure~\ref{fig:Mpop} shows the ``UniformQ" (top) and ``Bimodal" (bottom) primary mass distributions, inferred from a simulated
population of $50$ observations with realistic measurement uncertainties. Both populations have a sharp upper limit (pink histograms).
The lack of observations with masses above that value -especially
since they are favored by selection effects- results in a similarly sharp cutoff in the inferred distribution (green shaded regions).
The minimum mass does not result in a sharp cutoff of the $m_1$ distribution as $m_2<m_1$, but in a gradual decline. This
 decline together with the inferred mass ratio distribution results in a measurement of $\Mmin$.


\section{Results}
\label{sec:results}

The chirp mass $\Mchirp$ is the best measured intrinsic parameter for all binaries observed to date~\cite{2018arXiv181112907T} and largely drives mass inference,
especially for low-mass systems. Its inferred value provides a sharp cutoff for both the maximum and the minimum mass possible for the binary components
of $2^{1/5} \Mchirp $. For example, from the inferred chirp mass alone we know that GW170817 contains an object with mass
 $\lesssim 1.36\Msolar$~\cite{Abbott:2018wiz}, while GW190425 has an object with a mass $\gtrsim1.65\Msolar$~\cite{Abbott:2020uma}, providing
 some first crude bounds on $\Mmin$ and $\Mmax$ from GWs. The same applies to our simulated populations, where we expect sharp upper and
 lower limits on $\Mmin$ and $\Mmax$ based on the smallest and largest observed $\Mchirp$ respectively.

\begin{figure}[h]
\includegraphics[width=\columnwidth,clip=true]{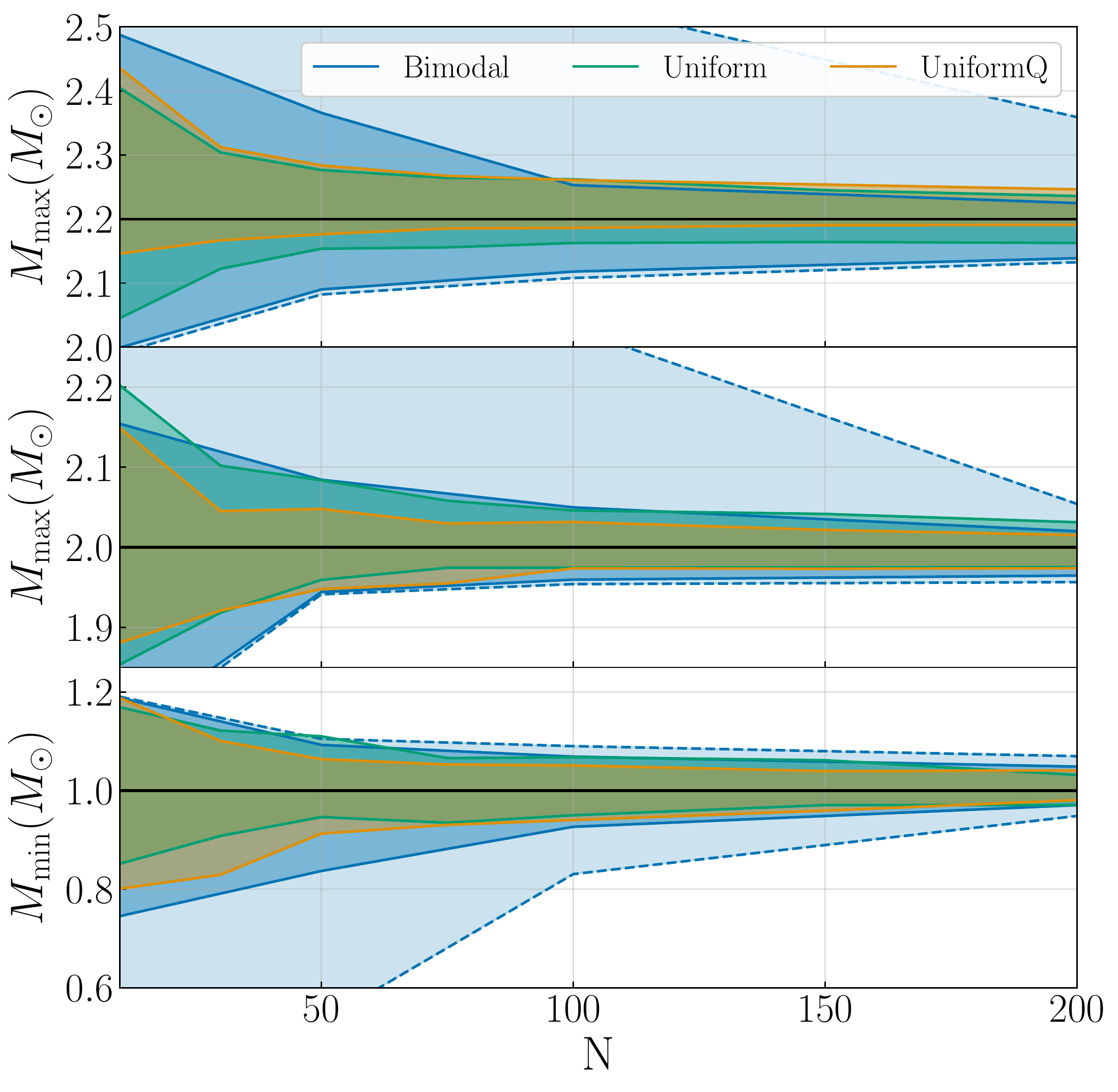}
\caption{Highest probability credible intervals on the maximum (top) and the minimum mass
(bottom) as a function of the number of detections, averaged over 20 populations.
Solid horizontal lines denote the true value.
For the ``Uniform" and ``UniformQ" distributions we show 90\% intervals. For the ``Bimodal" distribution we show
70\% (dark) and 90\% (light shading) intervals.  The intervals are asymmetric because the maximum and minimum masses must enclose all observed events, but are typically less constrained away from the observed masses.
}
\label{fig:Mconv}
\end{figure}

 Figure~\ref{fig:Mconv} shows the expected measurement uncertainty for $\Mmin$ and $\Mmax$ as a function of the number of observed signals $N$ for different
 mass distributions averaged over population realizations. Shaded regions correspond to highest probability density intervals, though the $\Mmin$ and $\Mmax$
 posteriors are fairly asymmetric due to the effect described above. In all cases we find that we can extract the correct values, as expected for inference where the
 model matches the intrinsic distribution of sources.

We find that if NS masses are uniformly distributed (green and orange) $30-50$ signals, potentially detectable during the fourth observing run circa 2022~\cite{Aasi:2013wya}, can lead to an
  estimate of $\Mmin$ to within $\sim 0.2M_{\odot}$ and $\Mmax$ within $\sim 0.1 M_{\odot}$ at the 90\% level. Further detector improvements and new observatories can potentially lead to the detection of $100-200$ systems, though estimates are
  uncertain~\cite{Aasi:2013wya}. With $200$ detections we can extract $\Mmin$ to within $\sim 0.1 M_{\odot}$
  and $\Mmax$ within $\sim 0.05 M_{\odot}$ at the 90\% level using GW mass inference alone.

If the NSs observed by GW detectors follow a bimodal
distribution instead, then we expect fewer NSs
close to the maximum and minimum, and all constraints are correspondingly
weaker. This could, for example, be the case if binary formation 
disfavors systems containing the heaviest or lightest possible NSs;
even in this case, though, any sharp upper cutoff in the mass distribution is the result of the nuclear EoS. 
Our assumed bimodal mass distribution has
$(\mu_1,\sigma_1,\mu_2,\sigma_2)=(1.35,0.07,1.85,0.35)\Msolar$ and $A=0.63$,
implying $0.12N(0.05N)$ systems with $m_1>2.0(2.2)\Msolar$; these parameters are
a ``fair draw'' from the posterior over NS mass distributions fitted to galactic pulsars
 \cite{Alsing:2017bbc}.  Our mass model ``learns'' the maximum (minimum)
NS mass from the absence of observed events above (below) the cutoff mass.  Such
an absence can only be inferred confidently when the corresponding smooth mass
distribution without the cutoff would have produced several events above (below)
the cutoff. If the smooth distribution predicts five ``missing systems'' above (below) the cutoff,
the probability of observing none is smaller
than 1\%, and the existence of a cutoff can be confidently inferred; for the
bimodal mass distribution, $N\sim50(100)$ detections would
yield $\sim 5$ detections above $\Mmax=2.0(2.2)\Msolar$.  We therefore expect
that 50--100 detections from the bimodal mass distribution are required to
confidently identify the cutoff mass scale.   This expectation is confirmed by
 Fig.~\ref{fig:Mconv} where we plot both the 70\%
(dark) and 90\% (light shading) credible interval on the cutoff masses. The
posteriors for $\Mmax$ and $\Mmin$ are highly asymmetric because the cutoff must
always be larger/smaller than the heaviest/lightest observation, but
generally less constrained in the opposite direction.

The above estimates assume no a priori restrictions on the NS spins; instead
assuming that merging NSs have low spins would result in tighter inference of
all parameters by mitigating the spin -mass ratio
correlation~\cite{Abbott:2018wiz}.

\section{Discussion}
\label{sec:disc}

A robust determination of the maximum and minimum NS mass can have implications for our understanding of the high-density EoS of NSs.
The ever increasing lower bound on the maximum mass driven from pulsar observations has been used to rule out the softest EoS models, leading
to the current picture of EoSs predicting almost constant radii for NSs in the range $1-1.8\Msolar$~\cite{Ozel:2016oaf}. An upper limit on the maximum mass
should lead to a complementary constraint on the stiffness of the EoSs.

\begin{figure}[h]
\includegraphics[width=\columnwidth,clip=true]{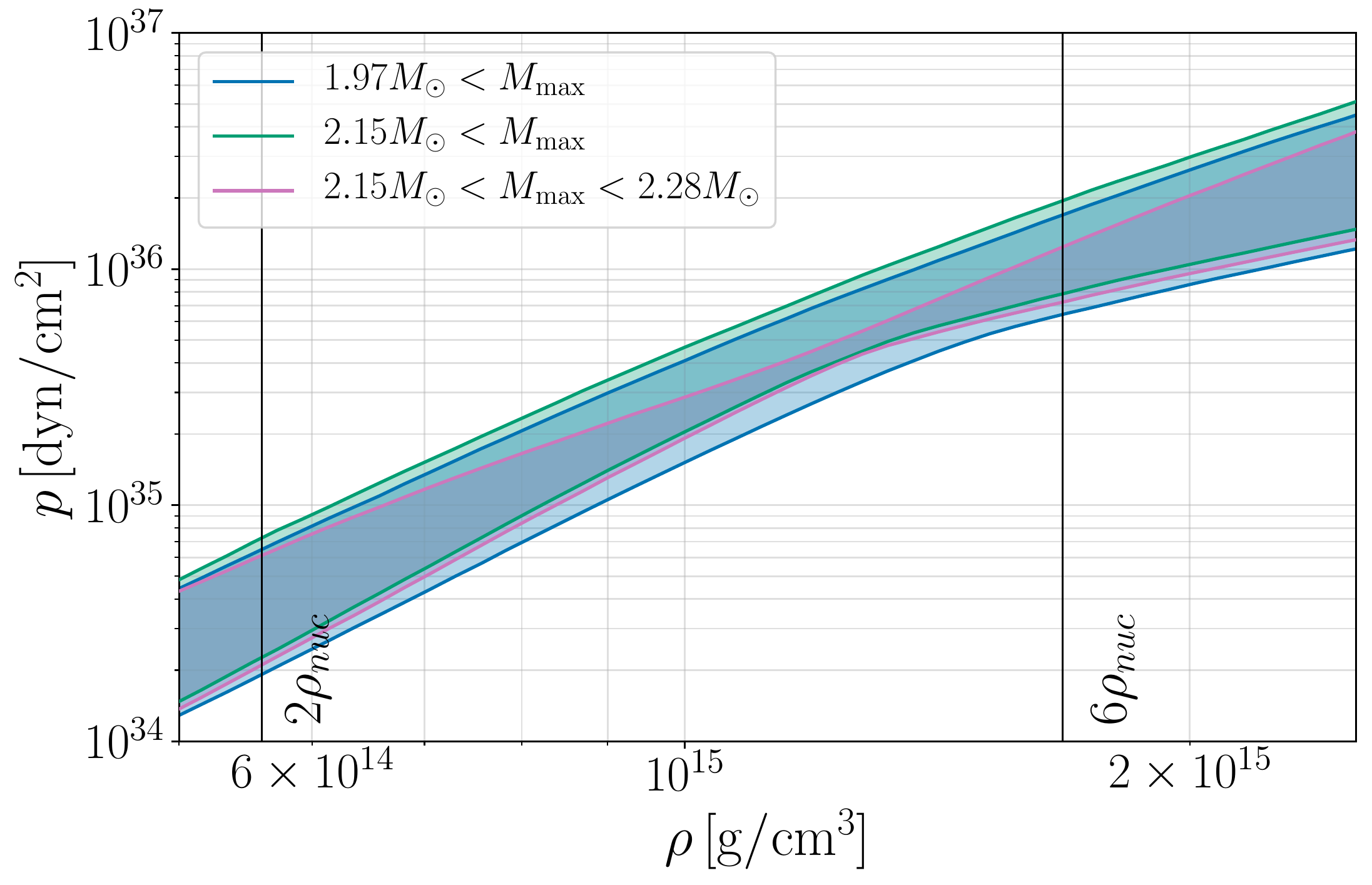}
\caption{Effect of a maximum mass constraint on EoS inference. We plot 90\% credible intervals for the NS pressure as a function of the density for GW170817
under different potential maximum mass knowledge. As expected, the maximum NS mass offers information about the high-density EoS, around $4-5$ times
the nuclear saturation density~\cite{}.
}
\label{fig:Prho}
\end{figure}

Figure~\ref{fig:Prho} sketches the effect of potential maximum mass
constraints on EoS inference for GW170817~\cite{Abbott:2018exr,170817samples,Carney:2018sdv}. The blue band corresponds to current constraints that already
assume that the maximum mass is above $1.97\Msolar$~\cite{Antoniadis:2013pzd}. Incorporating new $\Mmax$ constraints would preferably make
use of the full inferred distribution~\cite{Miller:2019nzo} while avoiding biases caused by mishandled Occam
penalties~\cite{Landry:2020vaw}. However, we can make a quick
estimate of the effect of $\Mmax$ on EoS inference by imposing an upper and a lower limit on $\Mmax$ corresponding to its
90\% interval after the detection of $\sim 50$ BNS signals with
the ``UniformQ" distribution. The green
shaded band is the result of an even more stringent lower limit on $\Mmax$ and it rules out some of the soft parameter space at pressures around 4-5 times the
nuclear saturation density. Adding an upper limit on $\Mmax$ leads to the pink shaded region which additionally constrains the stiff part of the EoS at similar
densities.

Overall, a $\sim 6$\% constraint on $\Mmax$ leads to a constraint of the pressure at 4.5 times the saturation density
of $\sim 20$\%, possible with $\sim50$ detections. The fact that the pressure constraint is stronger in this density region is due to the fact that
the maximum mass is correlated with the high-density EoS~\cite{Ozel:2016oaf}.
Such constraints on the high-density EoS might only be achievable through measurements of $\Mmax$ in the near future. Tidal measurements of binaries with masses
close to maximum are intrinsically challenging as tidal interactions are weaker for more massive NSs~\cite{Flanagan:2007ix,Hinderer:2007mb,Hinderer:2009ca}.
Further GW probes of high densities such as post merger emission from a hyper massive remnant are expected to be detected on longer timescales than the first
$50$ BNS signals~\cite{Torres-Rivas:2018svp}. Finally, an accurate measurement of $\Mmax$ can be compared to tidal inference from the BNS inspiral which probes
the low-density EoS to potentially probe signatures of a phase transition in the EoS~\cite{Bauswein2013,Han:2018mtj,Montana:2018bkb,Chatziioannou:2019yko}.

On the astrophysical side, a determination of $\Mmin$ could inform the boundary between WDs and NSs. Ground based GW detectors are deaf to signals
 from binaries containing WDs; the orbital separation of a binary emitting at $10$Hz -a common lower boundary on the LIGO bandwidth- is $\sim 600$km
for a total mass of $2\Msolar$. Any binary containing WDs would merge below $10$Hz and so any inspiral signal seen
in LIGO must contain objects more compact than WDs. Determination of $\Mmin$ 
could aid the classification of low-mass binary components and inform about their formation~\cite{Tauris:2019sho}.

Besides WDs, BHs that form through stellar evolution are expected to be heavier than NSs and not lead to a contamination of the BNS population. However, exotic
possibilities such as primordial BHs or merger products could occupy any mass range, even below $1\Msolar$. Searches for subsolar mass binaries through GWs
place upper limits on their abundance~\cite{Magee:2018opb,Abbott:2018oah,Authors:2019qbw} however these are less stringent than the upper limit of
the inferred BNS rate~\cite{Abbott:2020uma} due to the decreased detector sensitivity to low mass signals. Such BHs could be differentiated from NSs
by the fact that the latter are expected to be subjected to strong tidal effects, while the former do not~\cite{Chen:2019aiw}. 

On the high mass side, the existence of
BHs with masses comparable to the most massive NSs would alter the mass distribution of low-mass objects, and possibly fill in the low mass gap. 
If low-mass BHs are, as expected, less abundant than NS
of similar mass, then the mass distribution would no longer terminate at $\Mmax$, but it would exhibit a sharp drop reaching either zero if there is a gap between
NSs and BH, or a finite value if there is no such gap.
In either case, though, the sharp drop is caused by the maximum NS mass and it can be detected with similar methods and comparable accuracy as estimated here.
Indeed, such a combined analysis of the mass distribution of the O2 detections was recently presented in~\cite{Fishbach:2020ryj}, where it was argued that
there is tentative evidence for a non trivial feature in the mass distribution between NSs and BHs in the form of a gap.
If, on the other hand, BHs outnumber NSs in the
$2-3\Msolar$ range (a scenario that is observationally disfavored~\cite{Fishbach:2020ryj}), then telling them apart will be very challenging, requiring electromagnetic observations~\cite{Barbieri:2020ebt}
or next generation detectors that could constrain
the tidal signature of $\sim 2\Msolar$ compact objects~\cite{Chen:2020fzm}.

A sharp feature in the NS mass distribution could in principle break the
degeneracy between distance and redshift in GW observations and lead to
constraints on the Hubble constant $H_0$ \cite{Taylor:2012PhRvD..85b3535T}; a
similar approach has been proposed for BHs \cite{Farr:2019twy}. However, for a
typical distance uncertainty of $50\%$~\cite{Farr:2015lna} and an
uncertainty of $0.1\Msolar$ in the cutoff mass, local ($z \lesssim 0.1$) signals
will not provide sufficient accuracy in the redshift measurement to permit
measurement of $H_0$ comparable with competing constraints~\cite{Chen:2017rfc}.
With third-generation GW detectors~\cite{2011CQGra..28i4013H,Reitze:2019iox} the
 reach for neutron star systems extends to sufficiently high redshift
that a $0.1\Msolar$ mass uncertainty would be sufficient to determine the
redshift-distance relation at the subpercent level
\cite{Taylor:2012PhRvD..86b3502T}, but by the time such detectors are operating
other GW methods \cite{Chen:2017rfc} will likely have already
achieved subpercent accuracy in $H_0$.

Finally, we argue that the determination of $\Mmax$ directly from the NS mass distribution is expected to be less prone to common systematic uncertainties. Mass
measurement for BNS signals is driven by the low-order terms in the phase evolution which are well understood and modeled. Complementary methods of
inferring the EoS and maximum mass simultaneously~\cite{Landry:2018prl,Essick:2019ldf,Wysocki:2020myz} rely on accurate tidal inference with improved waveform
models than currently available and modeling of the EoS itself to extrapolate from low to high densities. At the same time, methods based on
information about the fate of the merger remnant are subject to systematics related to the interpretation of the post merger evolution and the
electromagnetic emission~\cite{Margalit:2017dij,2018ApJ...852L..25R,Margalit:2019dpi}.
In practice, we anticipate a multitude of methods utilizing different assumptions to be employed on future data;
both a potential agreement and a potential disagreement between
the different methods will teach us something about NSs and their properties.

\section*{Acknowledgments}

We thank Hsin-Yu Chen, Paul Lasky, Cole Miller, and Eric Thrane for useful discussions. We thank Phil Landry and Bernard Whiting for carefully reading the manuscript. The Flatiron Institute is supported by the Simons Foundation. 
Software: {\tt matplotlib}~\cite{Hunter:2007}, {\tt stan}~\cite{JSSv076i01},
{\tt numpy}~\cite{numpy}, {\tt scipy}~\cite{2020SciPy-NMeth}, {\tt astropy}~\citep{astropy:2013, astropy:2018}.

\appendix

\section{Population Simulation}
\label{sim}

The posterior distributions for the source parameters of observed GW signals are typically computed through stochastic
sampling methods~\cite{Veitch:2014wba,TheLIGOScientific:2016wfe}. For this study we consider hundreds of simulated BNS signals, which would make
stochastic sampling from the full multidimensional posterior distribution computationally prohibitive. In this appendix, we instead describe how we
estimate the measurement uncertainty for our simulated signals.

We draw parameters for each simulated system from a relevant astrophysical distribution. We assume that the SNR $\rho$ is distributed according to
$\rho^{-4}$~\cite{Chen:2014yla}, a reasonable assumptions for noncosmological sources such as BNSs detected with detectors in current sensitivity.
Though our analysis only considers the mass and not the spin distribution of BNSs, mass and spin measurements are correlated. We therefore simulate both in
our population in order to achieve realistic mass measurement uncertainties.
The effective spin $\chieff$ (see Appendix~\ref{phase}) is assumed to be uniformly
distributed in $[-0.05,0.05]$, while the mass distributions we consider (``Uniform", ``UniformQ", and ``Bimodal") are described in the main text.

Given the true parameters of the system, we approximate the likelihood for each parameter based on the following considerations.
The main observable from BNS signals is the
GW phase, whose evolution is determined by the system parameters. For long inspiral signals the phase can be expressed as a Taylor
expansion around small velocities or, equivalently, large separations.
This post-Newtonian (PN) expansion introduces terms at each order that depend on the system parameters. The
first three terms in the expansion encode the component masses and spins with
the PN coefficients $\Psi_0, \Psi_2$ and $\Psi_3$ corresponding to 0PN, 1PN, and 1.5PN orders respectively (a term of
$N$PN order contains an extra factor of $(u/c)^{N/2}$ compared to the leading order term, where $u$ is a characteristic velocity of the system, and $c$ is the speed of light).
Their form is given in Appendix~\ref{phase}.

Given this, instead of assuming that the likelihood is gaussian in the parameters of interest (the component masses and spin) as is common, we assume that it is
gaussian in $\Psi_0, \Psi_2$ and $\Psi_3$ with a standard deviation of $\sigma_0$, $\sigma_2$, and $\sigma_3$ respectively~\cite{Ng:2018neg}.
Effectively, this amounts to a gaussian approximation for the likelihood (ie. a Fisher matrix-based analysis), but where the gaussian assumption is applied to
$\Psi_0, \Psi_2$ and $\Psi_3$ rather than the binary parameters.
The values of the standard deviations are conservatively determined by comparison to the high-spin available results for
GW170817~\cite{170817samples} and GW190425~\cite{190425samples}:
$\sigma_0=0.0046 \Psi_0/\rho$, $\sigma_2=0.2341\Psi_2/\rho$, and $\sigma_3=-0.1293 \Psi_3/\rho$, where we have also assumed that each measurement uncertainty is inversely
proportional to the signal SNR.

For each binary with true parameters $(m_1,m_2,\chieff)$ and an SNR $\rho$ we compute $\Psi_0,\Psi_2,\Psi_3$.  We then draw $\Psi_{i,\mathrm{obs}}$ from ${\cal{N}}\left( \Psi_i, \sigma_i \right)$.  The likelihood for each PN term
$\Psi_i$ is ${\cal{N}}\left(\Psi_{i,\mathrm{obs}},\sigma_i\right)$, i.e. a normal distribution with a standard deviation of $\sigma_i$ centered at the ``observed'' $\Psi_{i,\mathrm{obs}}$.
We then sample independently from the likelihoods for the three PN coefficients and transform the result into samples for the likelihoods of
$m_1, m_2, \chieff$, taking into account the appropriate transformation Jacobian.

\section{Gravitational wave phase}
\label{phase}

In this appendix for completeness we collect the GW phase terms we use in order to simulate our BNS populations.
Consider a compact binary with component masses $m_1$ and $m_2$ with $m_1>m_2$ and dimensionless spins $\chi_1$ and $\chi_2$. In the following,
we ignore the effect of spin precession~\cite{Apostolatos:1994mx}, as NS spins are expected to be small and there is
no evidence for precession in the two detected BNS signals~\cite{Abbott:2018wiz,Abbott:2020uma}.
This is a conservative assumption as spin precession could potentially improve the measurement of the binary masses
and spins~\cite{Hannam:2013uu,Chatziioannou:2014coa}. We define
$\Mchirp = (m_1 m_2)^{3/5}/(m_1+m_2)^{1/5}$, the chirp mass, $q=m_2/m_1$, the mass ratio, $\nu =q/(1+q)^2$, the symmetric mass ratio,
$\delta m = (m_1-m_2)/(m_1+m_2)$, the mass difference, $\chieff = (m_1 \chi_1+m_2 \chi_2)/(m_1+m_2)$, the effective spin, and $\chi_a=(\chi_1-\chi_2)/2$,
the spin difference.

The phase of the frequency domain GW signal up to 1.5PN under the stationary phase approximation~\cite{Droz:1999qx} is given by~\cite{Blanchet:2013haa}
\begin{align}
\Psi(f) &= 2 \pi f t_c -\phi_c-\frac{\pi}{4}\nn \\
&+\Psi_0(\Mchirp)f^{-5/3}+\Psi_2(\Mchirp,\nu)f^{-1}+\Psi_3(\Mchirp,\nu,\beta)f^{-2/3},
\end{align}
where $t_c$ is the time of coalescence, $\phi_c$ is the phase of coalescence, and the three terms in the second line are the 0PN, 1PN, and 1.5PN terms respectively.
The coefficient of each term is a function of the system intrinsic parameters with
\begin{align}
\Psi_0(\Mchirp) &= \frac{3}{128 \Mchirp^{5/3} \pi^{5/3}},\\
\Psi_2(\Mchirp,\nu) &= \frac{5}{96 \Mchirp \pi \nu^{2/5}}\left(\frac{743}{336}+\frac{11\nu}{4} \right),\\
\Psi_3(\Mchirp,\nu,\beta) &=\frac{3 \left(4\beta-16\pi\right)}{128 \Mchirp^{2/3} \pi^{2/3}\nu^{3/5}},
\end{align}
where $\beta$ is a linear function of the spins, encoding the leading-order spin-orbit coupling. The leading-order 0PN term, $\Psi_0$,
is a function of the chirp mass only;
being the largest contribution to the GW phase, this term is measured to exquisite precision for BNSs, which have typical $\Mchirp$ measurement errors of
${\cal{O}}(10^{-4})$~\cite{PhysRevD.49.2658}.
The 1PN term, $\Psi_2$, depends on the ratio of the binary component masses, and can be used in conjunction with $\Mchirp$ to measure
the individual masses~\cite{1976ApJ...210..764W}.
The 1.5PN coefficient, $\Psi_3$ contains two terms of different origin. The second term in the parentheses, proportional to $16\pi$, is a
so-called tail term~\cite{PhysRevD.47.1497}, arising from scattering of the GWs off of the spacetime curvature as they propagate outwards from the binary near zone.
The first term in the parentheses, proportional to $\beta$, arises from the spin-orbit interaction between the binary components~\cite{PhysRevD.47.R4183}, given by
\begin{equation}
\beta = \frac{1}{3}\left(\frac{113-76\nu}{4}\chieff+\frac{76}{4}\delta m \nu \chi_a\right).
\end{equation}

The simultaneous presence of $\beta$ and $\nu$ in $\Psi_3$ results in the infamous spin-orbit degeneracy, deteriorating the measurement of both mass ratio and
spins from GW signals~\cite{PhysRevD.49.2658}.
Additionally, $\beta$ represents the leading-order spin contribution, and it is thus the best measured spin parameter, akin to the chirp mass.
It is common to disregard the second term in $\beta$ and study directly the effective spin $\chieff$ for two reasons: (i) the second term is proportional to the mass
difference and could be small, especially for BNS systems, and (ii) the effective spin $\chieff$ is conserved to at least 2PN order under spin precession and radiation
reaction~\cite{Racine:2008qv}. We do the same for our simulations.

\bibliography{OurRefs}

\end{document}